# Quasi-particle interference and superconducting gap in a high-temperature superconductor $Ca_{2-x}Na_xCuO_2Cl_2$

T. Hanaguri[1,2*], Y. Kohsaka[3], J. C. Davis[3], C. Lupien[4], I. Yamada[5], M. Azuma[5], M. Takano[5], K. Ohishi[6], M. Ono[1,2] & H. Takagi[1,2,7]

1.  Magnetic Materials Laboratory, RIKEN (The Institute of Physical and Chemical Research), Wako 351-0198, JAPAN.
2.  CREST, Japan Science and Technology Agency, Kawaguchi 332-0012, JAPAN.
3.  LASSP, Department of Physics, Cornell University, Ithaca NY 14853 USA.
4.  Département de Physique, Université de Sherbrooke, Sherbrooke, QC J1K 2R1 CANADA.
5.  Institute for Chemical Research, Kyoto University, Uji 601-0011, JAPAN.
6.  Advanced Science Research Center, Japan Atomic Energy Agency, Ibaraki 319-1195, JAPAN.
7.  Department of Advanced Materials, University of Tokyo, Kashiwa 277-8561, JAPAN.
*   e-mail: hanaguri@riken.jp

**High-transition-temperature (high-$T_c$) superconductivity is ubiquitous in the cuprates containing $CuO_2$ planes but each cuprate has its own character. The study of the material dependence of the $d$-wave superconducting gap (SG) should provide important insights into the mechanism of high-$T_c$. However, because of the 'pseudogap' phenomenon, it is often unclear whether the energy gaps observed by spectroscopic techniques really represent the SG. Here, we report spectroscopic imaging scanning tunneling microscopy (SI-STM) studies of nearly-optimally-doped $Ca_{2-x}Na_xCuO_2Cl_2$ (Na-CCOC) with $T_c = 25 \sim 28$ K. They enable us to observe the quasi-particle interference (QPI) effect in this material, through which unambiguous new information on the SG is obtained. The analysis of QPI in Na-CCOC reveals that the SG dispersion near the gap node is almost identical to that of $Bi_2Sr_2CaCu_2O_y$ (Bi2212) at the same doping level, while $T_c$ of Bi2212 is 3 times higher than that of Na-CCOC. We also find that SG in Na-CCOC is confined in narrower energy and momentum ranges than Bi2212. This explains at least in part the remarkable material dependence of $T_c$.**

Recent progress in spectroscopic techniques has enabled the elucidation of various interesting aspects of the electronic states of high-$T_c$ cuprates in real and momentum ($\mathbf{k}$) spaces. Scanning tunneling microscopy/spectroscopy (STM/STS) measurements have provided a wide variety of real-space information[1] including various types of electronic inhomogeneity[2-5] and the presence of energy non-dispersive 'checkerboard' (CB) modulations in the tunneling-conductance map $g(\mathbf{r},E) \equiv dI/dV(\mathbf{r},E)$ (refs. 6-9). In $\mathbf{k}$ space, angle-resolved photoemission spectroscopy (ARPES) on the underdoped cuprates has revealed that coherent quasi-particle states exist only in the nodal region around ($\pi/2$, $\pi/2$) (refs. 10-12). Here, the ARPES spectra away from the node are very broad. As a result, the cylindrical Fermi surface (FS) centred around ($\pi$, $\pi$) is somehow 'truncated' around the 1st Brillouin zone face and ends up with a so-called 'Fermi arc' in $\mathbf{k}$ space[13]. The Fermi arc length increases with doping and the full FS is restored in the optimally/overdoped region[12].

It is very important to understand how superconductivity emerges from such a complicated situation. Because superconducting states are characterised by the SG, detailed information on the SG is indispensable. There is now consensus that the SG in cuprates has $d$-wave symmetry[14] and thus has nodes in $\mathbf{k}$ space. However, direct studies of the SG structure by spectroscopic techniques are still in a pioneering stage[1,10,15,16]. The reason is that, while energy gaps can be measured in cuprates, whether such gaps represent the true SG is often unclear because of the presence of the 'pseudogap'[17].

Here we choose SI-STM as a tool to settle this issue. Of course, because of the electronic heterogeneity, the atomic spatial resolution of SI-STM is highly advantageous. Perhaps more importantly, by using Fourier transformations, one can now access the $\mathbf{k}$-space electronic structure as well[9,18,19]. So far, numerous STM/STS studies have been performed on Bi2212. Among them, the discovery of dispersive $g$-modulations due to QPI in the $d$-wave superconducting state[9,18-22] and its use for simultaneous determination of the $\mathbf{k}$-space region supporting the $d$-wave quasi-particles and the dispersion of the SG, provided a completely new approach to studies of high-$T_c$ superconductivity[9,18,19]. Conventional spectroscopies essentially give us the density-of-state (DOS) spectrum only and can not distinguish the SG from the pseudogap. By contrast, when observed QPI pattern is fully consistent within the 'octet' model[19], one can be certain that the data represent the true SG. (See the next section.) In spite of this great advantage, detailed information from QPI has been available only in Bi2212 (refs. 9,18,19); the universal characterization of the cuprate SG using the QPI technique has not yet been achieved.

Here we report the QPI effect in nearly-optimally-doped Na-CCOC with $T_c = 25 \sim 28$ K. Although the CB $g$-modulation dominates the STS data[8], the full octet model of QPI is confirmed by utilizing the contrast of spatial-phase relations between CB $g$-modulation and QPI. The SG is identified only on an 'arc' in $\mathbf{k}$ space near the gap node. No well-defined coherence peak is observed in the tunneling spectrum and the antinodal region therefore remains incoherent. Thus, Na-CCOC exhibits the apparent coexistence of nodal superconductivity with the incoherent antinodal states as in the case of underdoped Bi2212 (ref. 9), indicating that these are universal features of high-$T_c$ superconductivity. The SG dispersion of nearly-optimally-doped Na-CCOC is almost identical to that of optimally-doped Bi2212 with $T_c = 86$ K (ref. 19) but is confined in narrower energy and momentum ranges. From these observations we hypothesize that the energy scale which determines the $T_c$ of optimally-doped cuprates is not set by the superconducting gap dispersion alone but by the energy and momentum-space location where this gap terminates on the Fermi arc.

## QPI AND THE OCTET MODEL

Originally, QPI has been observed on conventional metal surfaces[23-25]. In the presence of elastic scattering, quantum interference of scattered electrons results in real-space electronic standing waves, characterized by wave vectors $\mathbf{q}$ connecting points on the contours of constant quasi-particle energy (CCE) in $\mathbf{k}$ space. This standing wave can be detected using SI-STM by mapping $g(\mathbf{r}, E)$, which reflects the local DOS at given location $\mathbf{r}$ and energy $E$. The wave vector $\mathbf{q}$ changes with $E$ according to the electron-energy-dispersion relation. The Fourier amplitude of the conductance map, $\left| g(\mathbf{q}, E) \right|$, allows us to determine characteristic $\mathbf{q}$ vectors at a given $E$. Therefore, the energy-dispersion relation can be obtained experimentally.

In the superconducting state, the Bardeen-Cooper-Schrieffer theory of superconductivity tells us that elementary excitations are not bare electrons or holes but coherent admixtures of particle and hole states known as the Bogoliubov quasi-particle[26]. The excitation energy of a Bogoliubov quasi-particle $E(\mathbf{k})$ is given by $E(\mathbf{k}) = \sqrt{\varepsilon(\mathbf{k})^2 + \Delta(\mathbf{k})^2}$ where $\varepsilon(\mathbf{k})$ is the normal-state band dispersion and $\Delta(\mathbf{k})$ is the SG. In the case of high-$T_c$ cuprates, the SG has $d$-wave symmetry and thus vanishes in the $(0, 0)$ - $(\pm\pi, \pm\pi)$ directions (nodes). Because Bogoliubov quasi-particles can be excited by either positive or negative energies and the Fermi velocity is much larger than the nodal gap slope, the excitation spectrum around the node is a particle-hole symmetric thin Dirac cone with a banana-shaped CCE as illustrated in Figs. 1a and 1b. QPI of Bogoliubov quasi-particle occurs in this situation[9,18-22]. The intensities of the standing waves with $\mathbf{q}$'s connecting the ends of 'bananas' should become large, since the DOS, which is inversely proportional to the slope of dispersion, becomes a maximum at the energy-dependent octet ends of 'bananas'. There are 7 possible vectors $\mathbf{q}_i$ ($i = 1 \sim 7$) from one of the octet ends as shown in Fig 1b. In total, 32 different $\mathbf{q}$'s (16 of them are independent.) may be observed in $|g(\mathbf{q}, E)|$ (Fig. 1c). If these numerous $\mathbf{q}$'s are observed in $|g(\mathbf{q}, E)|$ and disperse with $E$ according to the above-mentioned 'octet' model[19] in an internally consistent fashion, the presence of coherent Bogoliubov quasi-particles and the underling $d$-wave SG can be established without doubt. By analysing the $E$ dependence of $\mathbf{q}_i$, one can deduce the locations of order elements in $\mathbf{k}$ space, namely FS position, as well as the dispersion relation of the SG, $\Delta(\mathbf{k})$ (ref. 19). The octet model explains the observed $|g(\mathbf{q}, E)|$ in Bi2212 quite well and detailed information about the SG has been obtained[9,19].

## SEARCH FOR QPI IN Na-CCOC

Bi2212 is the only material for which QPI has been detected so far. Only after identifying the SG in other cuprates by QPI, can the universal features of the superconducting electronic state of cuprates be elucidated. Even more importantly, a clue for the cause of the strong variations of $T_c$ from material to material may then be inferred. To carry out the search for QPI, we chose Na-CCOC[27,28] as the next likeliest cuprate. Since Na-CCOC crystals cleave well, clean and flat surfaces necessary for STM/STS can be obtained easily. Previous STM/STS works on Na-CCOC have been performed only in the heavily underdoped region and revealed nano-scale electronic inhomogeneity[29] and CB $g$-modulation[8]. However, no QPI signal was detected although the QPI signal may merely be too weak due to the weakness of superconductivity in the heavily underdoped regime.

Na doping into Na-CCOC requires a high pressure of the order of several GPa[27,28]. Since higher doping demands higher pressure during crystal growth, available samples were limited up to $x \sim 0.12$ with $T_c \sim 22$ K. Recent progress in crystal growth allows us to synthesise suitably-sized nearly-optimally-doped single crystals. In the present study, we measured crystals with $T_c \sim 25$ K and 28 K, which correspond to $x \sim 0.13$ and 0.14, respectively. Here we report SI-STM measurements on Na-CCOC performed with a newly-developed ultrahigh-vacuum (UHV) very-low-temperature STM at RIKEN. Samples were cleaved *in-situ* at 77 K under UHV and were immediately transferred to the STM unit which was kept at low temperature ($<$ 10 K). All the data were taken at a temperature $T \sim 0.4$ K.

It is important to notice that the crystal structure, the atomic species in the blocking layer between $CuO_2$ planes, the number of $CuO_2$ planes in the unit cell, and the $T_c$ at optimal doping are all different between Na-CCOC and Bi2212. Since these materials are so completely disparate in the physics/chemistry of the blocking layers through which tunneling occurs, common QPI phenomena must derive from the only common characteristic that is intrinsic to the electronic structure of the $CuO_2$ plane in which high-$T_c$ superconductivity takes place.

Figure 2a shows a typical topographic image of the sample with $T_c \sim 25$ K. All the spectroscopic imaging measurements reported in this paper have been simultaneously performed with similar atomic-resolution topographic images to maintain the position registry. The cleaved surfaces exhibit a perfect square lattice with superimposed inhomogeneity, which are similar to the topographic images in the heavily underdoped samples[8,29]. Overall similarities to the heavily underdoped samples are also seen in $g(\mathbf{r}, E)$ and an averaged tunneling spectrum; $g(\mathbf{r}, E)$ shows apparent CB modulations (Fig. 2b) while the averaged spectrum displays a V-shaped pseudogap below ~100 meV (Fig. 2c).

We next focus on spectroscopic features at low energies. As shown in Fig. 2d, inside the V-shaped pseudogap, we found another gap-like feature below ~10 meV. Thus, there are at least two energy scales in the system: ~100 meV pseudogap and ~10 meV low-energy gap. The existence of two energy scales is also pointed out by recent ARPES[15] and Raman spectroscopy[16]. We found that the low-energy gap tends to disappear homogeneously over the field of view under magnetic fields (up to 11 T) or at elevated temperatures (up to 20 K), suggesting that superconductivity plays a role for the low-energy gap formation. These observations motivated us to search for QPI to probe directly the SG dispersion in this low $T_c$ but nearly optimally-doped high-$T_c$ cuprate.

QPI patterns in Bi2212 have been observed in raw $g(\mathbf{r}, E)$ or $|g(\mathbf{q}, E)|$ (refs. 9,18,19). However, in the case of Na-CCOC ($T_c \sim 28$ K, $x \sim 0.14$), $g(\mathbf{r}, E)$ and $|g(\mathbf{q}, E)|$ are primarily governed by the apparent CB $g$-modulations as shown in Figs. 3a and 3b and clear signature of QPI is difficult to recognize. Thus, we need to find a new technique to extract QPI patterns which may be hidden behind the apparent CB $g$-modulations.

## SEPARATION OF QPI FROM THE 'CHECKERBOARD'

To separate QPI from the apparent CB $g$-modulations we consider the spatial-phase relations in the conductance modulations. A comparison between the positive and negative bias $g(\mathbf{r}, \pm E)$ within the low-energy gap shows that $g(\mathbf{r}, +E)$ and $g(\mathbf{r}, -E)$ are spatially quite similar and the local conductance maxima/minima occur predominantly at the same location as shown in Figs. 3a and 3b. In other words, the apparent CB $g$-modulations have the same wavelength and are in-phase across the Fermi energy. The QPI patterns should also have the same wavelength in $g(\mathbf{r},+E)$ and $g(\mathbf{r},-E)$ since $E(\mathbf{k})$ is expected to be particle-hole symmetric as shown in Fig. 1a. The spatial-phase relation for QPI is not known precisely but it is plausible that there is a phase difference between $g(\mathbf{r},+E)$ and $g(\mathbf{r},-E)$ as there is a sum-rule $\sum_n v_n^2(\mathbf{r})+u_n^2(\mathbf{r})=1$ between particle- and hole-like amplitudes of the Bogoliubov quasi-particle, $v_n^2(\mathbf{r})$ and $u_n^2(\mathbf{r})$, where $n$ is an appropriate quantum number[30]. Therefore, the QPI pattern may be extracted by means of the spatial-phase sensitive method if they have a non-trivial phase difference.

For this purpose, we propose to examine the ratio map defined by $Z(\mathbf{r},E)\equiv g(\mathbf{r},+E)\big/g(\mathbf{r},-E)$ and its Fourier transform $|Z(\mathbf{q},E)|$. Let us introduce a basic property of $|Z(\mathbf{q},E)|$. If the conductance modulation amplitude at finite $\mathbf{q}$, $g_q^{\pm}\equiv g(\mathbf{q},\pm E)$, is much smaller than the uniform component $g_0^{\pm}\equiv g(\mathbf{q}=0,\pm E)$, $|Z(\mathbf{q},E)|$ can be approximated as (to the first order)

$$|Z(\mathbf{q},E)|\approx\left(g_0^+\big/g_0^-\right)\left|\left(g_q^+\big/g_0^+\right)-\left(g_q^-\big/g_0^-\right)\right|.$$

In the present case, the relation $\left|g_q^{\pm}\right|\big/g_0^{\pm}<<1$ holds for all the energies studied. Therefore, the above approximation is well justified. Note that $|Z(\mathbf{q},E)|$ is sensitive to the relative signs between $g(\mathbf{q},+E)$ and $g(\mathbf{q},-E)$. (Note that $g_0^{\pm}>0$.) Namely, taking the ratio reduces the in-phase component.

In Figs. 3c and 3d, respectively we show the measured $Z(\mathbf{r},E=6\,\text{meV})$ and the Fourier transform $|Z(\mathbf{q},E=6\,\text{meV})|$. It is immediately evident that the CB $g$-modulation, which dominates $g(\mathbf{r},E)$ and $|g(\mathbf{q},E)|$, almost completely disappears. And, most satisfyingly, another type of modulation emerges. The features of the new modulations are clearly manifested in $|Z(\mathbf{q},E)|$ where numerous spots are observed at incommensurate $\mathbf{q}$'s. All of these $\mathbf{q}$'s are consistently explained in terms of the octet model[19] of QPI as shown by symbols in Fig. 3d. Thus the existence of the QPI effect and therefore the coherent Bogoliubov quasi-particle of the $d$-wave superconducting state are clearly confirmed for the first time in Na-CCOC.

We would like to note that the ratio $Z$ has another important feature besides its phase sensitivity[31]. In general, tunnelling conductance $g(\mathbf{r},E=eV)$ is given by

$$g(\mathbf{r}, eV) = \frac{eI_t N(\mathbf{r}, eV)}{\int_0^{eV_s} N(\mathbf{r}, E) dE}.$$

Here, $e$ is a unit charge, $I_t$ and $V_s$ are set-up current and bias voltage for feedback, respectively, and $N$ is a local DOS. Note that $g$ is not proportional to $N$ but is normalized by the integrated $N$ which is strongly influenced by the set-up condition through $V_s$. On the other hand, $Z(\mathbf{r}, eV)$ reflects the ratio of the $N$ precisely, because the denominator and pre-factor $eI_t$ are exactly cancelled out by taking the ratio[31]. This fact means that observed QPI is an intrinsic phenomenon free from any set-point related issues which inevitably contaminate $g(\mathbf{r}, E)$ and $|g(\mathbf{q}, E)|$. On the contrary, there is a possibility that CB $g$-modulation is affected by the set-point effect because it disappears in $Z$ at least in the energy range inside of the low energy gap where QPI is observed. Therefore, it is important to search for the possible electronic superstructure using the set-point-effect-free method.

Recently, ratio analyses at high energy (~150 meV) in underdoped Na-CCOC and Bi2212 have revealed the existence of an electronic-cluster glass (ECG) state characterised by the 4 Cu-Cu distance-wide bond-centered unidirectional electronic domains dispersed throughout without long-range order[31]. Similarities between ECG and CB $g$-modulation (e.g. characteristic length scale of 4 Cu-Cu distance) imply close relations between the two. It is interesting to note that apparently different phenomena, high-energy ECG and low-energy QPI, have been detected in the similar analysis scheme depending on the energy scale. Given the fact that the ratio analysis extracts intrinsic information on the electronic states, establishing the relation between ECG and QPI, which may represent localized and delocalized electronic states, respectively, represents a key goal for this field.

## INSIGHTS FROM THE QPI ANALYSIS

To analyze the observed Na-CCOC QPI quantitatively, we measured $\mathbf{q}_i$ at different $E$ and determined the energy-dependent locations of the octet ends of 'bananas' in $\mathbf{k}$ space. The obtained locus $\mathbf{k}(E) = (k_x(E), k_y(E))$ can be interpreted as a normal-state FS. We plot the obtained loci for two different samples in Fig. 4a. The result is consistent with a cylindrical FS centred around $(\pi, \pi)$ which is inferred from ARPES on underdoped samples[11].

We can determine the SG dispersion $\Delta(\mathbf{k})$ from $\mathbf{k}(E)$ as well[19]. In Fig. 4b, we plot the measured gap energy as a function of FS angle $\theta_\mathbf{k}$ about $(\pi, \pi)$. For comparison, SG dispersion of optimally-doped Bi2212 (ref. 19) is shown by a green line. Apparently, SG dispersions of these two compounds are quite similar. Since $T_c$'s are very different between Na-CCOC ($T_c \sim 28$ K) and Bi2212 ($T_c \sim 86$ K), we can conclude that the SG dispersion in the nodal region does not set the energy scale which determines $T_c$ alone.

It is worthy to note that QPI in optimally-doped Na-CCOC is observed in a rather limited part of the FS near the node ($25° < \theta_\mathbf{k} < 65°$). This suggests that available quasi-particle states are missing in the antinodal region[32-34] or that the antinodal quasi-particles may not have a correlation length long enough to produce interference due to disorder[35]. In accord with this, well-defined coherence peaks that manifest the coherent quasi-particle states in the antinodal region[9,19] are not observed in the tunneling spectra as shown in Figs. 2c and 2d. In the ARPES data on the $x \sim 0.12$ ($T_c \sim 22$ K) sample, with slightly lower hole concentration than the present samples, the spectral weight in the antinodal region is still substantially suppressed, which was pointed out to be linked with CB $g$-modulations[11]. Here we postulate that such incoherent antinodal states are not responsible for forming phase-coherent Cooper pairs. If this picture is correct, superconductivity is supported only by the coherent part of the FS around the node, namely, $d$-wave SG opens on the so-called Fermi arc. This conjecture predicts that the energy scale which sets $T_c$ is determined not only by the SG dispersion around the node but also by the effective 'arc length'. Analogous proposals have also been made based on the doping dependence of the gap amplitude[36].

Finally, we compare the tunneling spectrum of Na-CCOC with that of Bi2212. In optimally-doped Bi2212, well-defined coherence peaks are ubiquitous[9] and thus the entire FS may be responsible for superconductivity. On the other hand, in underdoped Bi2212, a different type of spectrum characterized by the V-shaped gap without a clear coherence peak appears, while QPI signal is still there[9]. This spectrum is very similar to the spectrum of Na-CCOC (refs. 8,29). Interestingly, in underdoped Bi2212, a non-dispersive CB $g$-modulation is also seen in the region where this type of spectrum is observed[9]. Therefore, it is reasonable to infer that the V-shaped gap without a clear coherence peak, the destruction of the antinodal state, and the CB $g$-modulation go hand in hand and are universal in high-$T_c$ cuprates. Another important common feature observed in both Na-CCOC and Bi2212 is that this state is superconducting with long-lived Bogoliubov quasi-particles which are manifested by QPI.

**CONCLUSION**

In summary we report the first observation of QPI in Na-CCOC. Using QPI studies of two different cuprates, we now begin to elucidate universal properties of the electronic structure of the $CuO_2$ plane. The full octet model of $d$-wave interference is confirmed in Na-CCOC; the SG and the arc upon which it exists in **k** space are determined. Interestingly, this nodal superconductivity coexists with the incoherent antinodal states. Comparison with Bi2212 data[9] suggests that this peculiar superconducting state is universal to high-$T_c$ cuprates. Furthermore, the SG dispersion from QPI in Na-CCOC is almost identical to that of Bi2212 at the same hole-doping[19]. Thus, the $T_c$ of an optimally-doped cuprate is not set by the SG dispersion alone but related to the coherence of the antinodal states.

**Acknowledgments**

The authors thank A. V. Balatsky, C. -M. Ho, D. -H. Lee, K. Machida, A. Mackenzie, K. McElroy, T. Tohyama, J. Zaanen and F. -C. Zhang for valuable discussions. They also thank J. Matsuno and P. Sharma for critical readings. T. H., M. T. and H. T. are supported by Grant-in-Aid for Scientific Research from the Ministry of Education, Culture, Sports, Science and Technology of Japan. J.C.D and Y. K. acknowledge support from Brookhaven National Laboratory under Contract No. DE-AC02-98CH1886 with the U.S. Department of Energy, from the U.S. Department of Energy Award# DE-FG02-06ER46306, and from the U.S. Office of Naval Research.

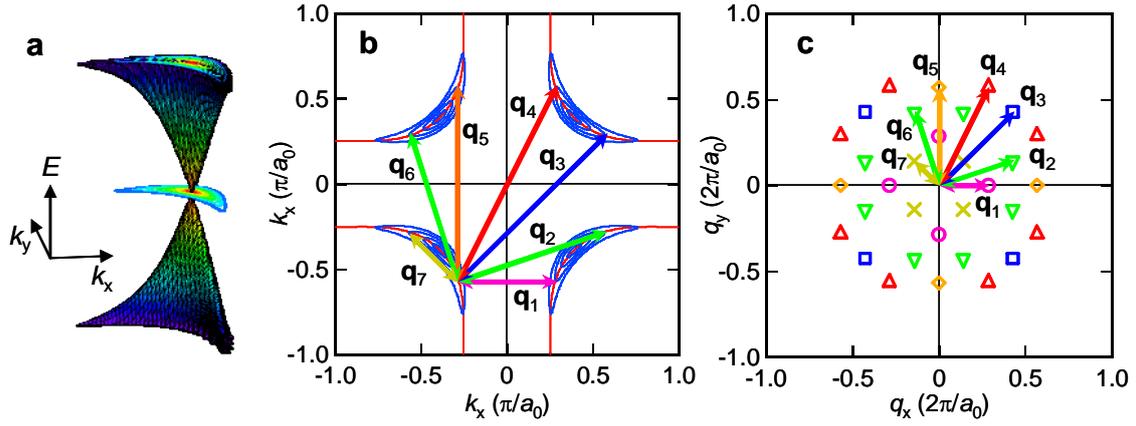

**Figure 1 Schematic illustration of electronic states of a high-$T_c$ superconductor responsible for the interference effect of Bogoliubov quasi-particles in the superconducting state. a**, A perspective view of the expected particle-hole symmetric quasi-particle dispersion relation $\pm\sqrt{\varepsilon(\mathbf{k})^2 + \Delta(\mathbf{k})^2}$ near one of the nodes showing banana-shaped CCE. **b**, Four sets of CCE (blue lines) in momentum space. The underlying FS is also shown by red lines. Note that the octet ends of 'bananas' tracks the FS. Since $\varepsilon(\mathbf{k})=0$ on the FS, the SG dispersion $\Delta(\mathbf{k})$ can be obtained by knowing the location of octet ends of 'bananas' in momentum space. Arrows denote expected scattering **q** vectors which characterize the standing waves of quasi-particles at a specific energy. Each vector connects the octet ends of 'bananas' where the density-of-states is high. Here, $a_0$ denotes the Cu-Cu distance. **c**, The **q**-space representation of the QPI, which is directly associated with the Fourier-transform of the conductance map $|g(\mathbf{q},E)|$. Symbols represent **q** vectors expected from the octet model[19]. The **q** vectors move with energy according to the FS shown in **b**. Thus FS can be reconstructed and $\Delta(\mathbf{k})$ can be determined from energy-dependent **q** vectors.

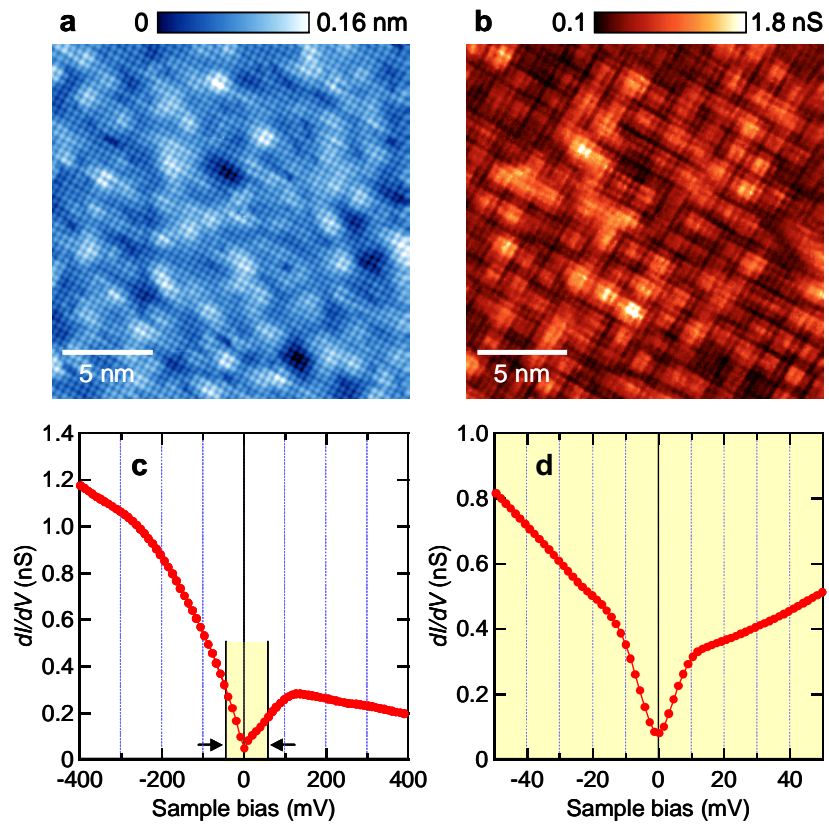

**Figure 2 Spectroscopic features of nearly-optimally doped Na-CCOC with $T_c \sim 25$ K ($x \sim 0.13$). a**, Typical topographic STM image of Na-CCOC. The image was taken at a sample-bias voltage $V_s = +400$ mV and a tunneling current $I_t = 100$ pA. All the spectroscopic images in this paper were taken simultaneously with similar atomic-resolution topographic images to maintain the atomic-resolution position registry. **b**, The conductance $g(\mathbf{r}, E)$ map at energy $E = +24$ meV in the same field of view. Apparent 'checkerboard' $dI/dV$-modulation, which has been found in heavily underdoped Na-CCOC (ref. 8), is observed even in the nearly-optimally doped sample. **c**, The spatially-averaged conductance spectrum showing the V-shaped pseudogap below ~100 meV with particle-hole asymmetry at higher energies. (Set-up condition: $V_s = +400$ mV and $I_t = 100$ pA.) **d**, The spatially-averaged spectrum at low energy region indicated by arrows in **c**, showing the opening of another energy gap inside of the pseudogap. (Set-up condition: $V_s = +150$ mV and $I_t = 100$ pA.) The central interest of this paper is on the electronic state in and around this low-energy gap.

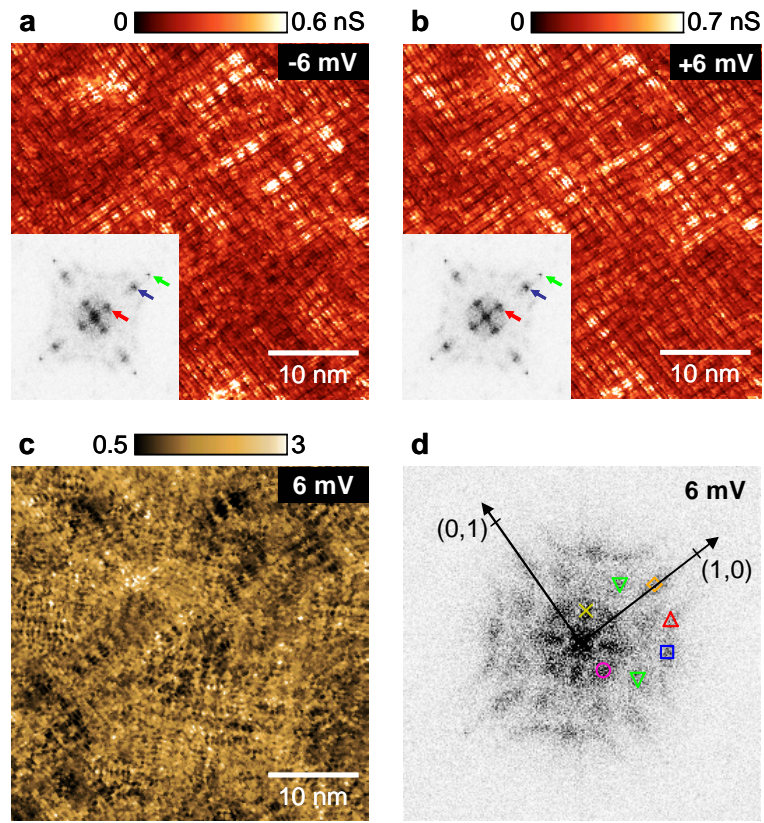

**Figure 3 Fourier-transform STS results on Na-CCOC ($T_c \sim$ 28 K, $x \sim$ 0.14) demonstrating the QPI in the conductance-ratio map** $Z(\mathbf{r}, E) \equiv \dfrac{g(\mathbf{r}, +E)}{g(\mathbf{r}, -E)}$. **a,** The filled-state $g(\mathbf{r}, -E)$ taken at -6 mV. (Set-up condition: $V_s$ = +150 mV and $I_t$ = 100 pA.) This map is basically dominated by the apparent CB $g$-modulation, which consists of components with wavelengths of $4a_0$, $4a_0/3$ and $a_0$ (ref. 8) as manifested by $|g(\mathbf{q}, E)|$ shown in the inset. (Red, dark blue and green arrows denote $4a_0$, $4a_0/3$ and $a_0$ components, respectively.) The same features are also seen in $g(\mathbf{r}, E)$ and $|g(\mathbf{q}, E)|$ in the empty state ($V_s$ = +6 mV) as shown in **b**. Note that $g(\mathbf{r}, E)$ (and $|g(\mathbf{q}, E)|$) are very similar between filled and empty states and local conductance maxima/minima occur predominantly at the same location. **c,** the conductance-ratio map $Z(\mathbf{r}, E)$ calculated from **a** and **b**. The CB $g$-modulation almost completely disappears and a new type of modulation emerges. **d,** $|Z(\mathbf{q}, E)|$ obtained by Fourier transformation of $Z(\mathbf{r}, E)$. The **q** vectors which characterize $Z(\mathbf{r}, E)$ are clearly represented by local maxima (dark spots) in the image. All of the **q** vectors expected from the octet model of QPI are observed in $|Z(\mathbf{q}, E)|$ as indicated by symbols which correspond to those in **Fig. 1c**. Note that no extra **q** vector is detected. This result strongly indicates that the modulation in $Z(\mathbf{r}, E)$ originates from the QPI.

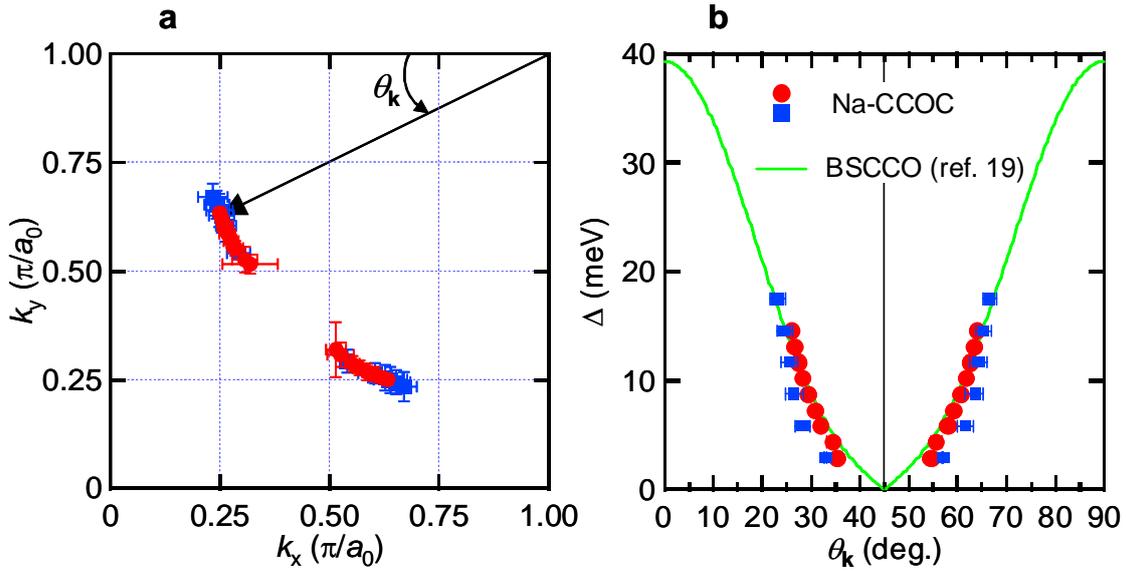

**Figure 4 The obtained FS and the SG dispersion for Na-CCOC ($T_c$ ~ 28 K, $x$ ~ 0.14). a**, The locus of octet ends of banana-shaped CCE in a quadrant of 1st Brillouin zone which represents the underlying FS. The data are symmetrized about the diagonal due to mirror symmetry. The results for two different samples are plotted in red circles and blue squares, respectively. In the analysis, we used 4 independent $\mathbf{q}_4(E) = (\pm 2k_x(E), 2k_y(E)), (2k_y(E), \pm 2k_x(E))$ which were clearly observed between 3 meV < $E$ < 15 meV. The other $\mathbf{q}_i$'s gave us consistent results but the observed energy range was narrower. The error bars represent the statistical scatters. **b**, The SG plotted as a function of the FS angle $\theta_k$ defined in **a**. The two different symbols have the same meaning as in **a**. The green line represents the SG dispersion for nearly-optimally doped Bi2212 with $T_c$ ~ 86 K (ref. 19). Even though $T_c$'s differ by a factor of three between these two compounds, the SG dispersions are very similar.